\documentclass[12pt,preprint]{aastex}
\begin{document}
\title{TrES-1: The Transiting Planet of a Bright K0V Star}
\author{
Roi Alonso\altaffilmark{1,2}, 
Timothy M. Brown\altaffilmark{2}, 
Guillermo Torres\altaffilmark{3},
David W. Latham\altaffilmark{3},
Alessandro Sozzetti\altaffilmark{4,3}, 
Georgi Mandushev\altaffilmark{5}, 
Juan A. Belmonte\altaffilmark{1},
David Charbonneau\altaffilmark{3,6},
Hans J. Deeg\altaffilmark{1},
Edward W. Dunham\altaffilmark{5}, 
Francis T. O'Donovan\altaffilmark{6},
Robert P. Stefanik\altaffilmark{3}
}
\altaffiltext{1}{Instituto de Astrof\'\i sica de Canarias, 38200 La Laguna,
Tenerife, SPAIN}
\email{ras@iac.es}
\altaffiltext{2}{High Altitude Observatory/National Center for Atmospheric
Research, 3450 Mitchell Lane, Boulder, CO 80307 USA}
\email{timbrown@hao.ucar.edu}
\altaffiltext{3}{Harvard-Smithsonian Center for Astrophysics, 60 Garden St.,
Cambridge, MA 02138 USA}
\altaffiltext{4}{University of Pittsburgh, Dept. of Physics \&
Astronomy, Pittsburgh, PA 15260 USA}
\altaffiltext{5}{Lowell Observatory, Flagstaff, AZ 86001, USA}
\altaffiltext{6}{California Institute of Technology, 1200 E. California Blvd.,
Pasadena, CA, 91125, USA}
\email{gtorres@cfa.harvard.edu}
\email{dlatham@cfa.harvard.edu}
\email{asozzett@cfa.harvard.edu}
\email{gmand@lowell.edu}
\email{jba@iac.es}
\email{dcharbonneau@cfa.harvard.edu}
\email{hdeeg@iac.es}
\email{dunham@lowell.edu}
\email{ftod@astro.caltech.edu}
\email{rstefanik@cfa.harvard.edu}

\begin{abstract}
We report the detection of a transiting Jupiter-sized planet orbiting a
relatively bright ($V=11.79$) K0V star.
We detected the transit light-curve signature in the course of the
TrES multi-site transiting planet survey, and confirmed the planetary
nature of the companion via multicolor photometry and precise radial
velocity measurements.  
We designate the planet TrES-1; its inferred mass is 0.75 $\pm
0.07\  M_{\rm Jup}$,
its radius is $1.08^{+0.18}_{-0.04}$ $R_{\rm Jup}$, and its orbital period is
3.030065 $\pm 0.000008$ days.
This planet has an orbital period similar to that of HD 209458b, but
about twice as long as those of the OGLE transiting planets.
Its mass is indistinguishable from that of HD 209458b, but its radius is
significantly smaller and fits the theoretical models without the need for
an additional source of heat deep in the atmosphere, as has been invoked
by some investigators for HD 209458b.
\end{abstract}

\keywords{binaries: eclipsing -- planetary systems -- stars: individual
(GSC 02652-01324) -- techniques: photometric -- techniques: radial velocities}

\section{Introduction}

Since before the discovery of the first transiting extrasolar planet
\citep{cha00, hen00}, 
it has been recognized that transits provide a sensitive way
to infer the existence of small bodies orbiting other stars \citep{str52}.
There are now dozens of ground-based photometric searches underway that aim
to detect planets of distant stars by means of their photometric
signatures \citep{hor03}, as well as several space projects with the same
purpose \citep{auv03, bor03}.

Until now, the only confirmed planet detections by transits
\citep{kon03, kon04, bou04} have been based on
the OGLE survey \citep{uda02a, uda02b, uda03}, 
which is performed with a telescope of 1.3m aperture.
The strategy of using a moderate-aperture telescope with seeing-limited
spatial resolution
must be commended for its obvious successes, and
moreover because the 3 OGLE planets are peculiar, having
the shortest orbital periods yet known.
But surveys using large telescopes suffer from 
the faintness of the stars with which
they deal (the $I$ magnitudes of the OGLE planet host stars range
from 14.4 to 15.7).
For such faint stars, the necessary follow-up observations are difficult
and time-consuming, and the precision with which planetary parameters
such as mass and radius can be determined is compromised.

For these reasons, we have pursued a transiting planet search organized
along different lines -- one that uses small-aperture, wide-field telescopes
to search for transits among brighter stars.
The principal challenges facing wide-field surveys such as ours
are to attain adequate photometric precision in the face of spatially
varying atmospheric extinction and instrumental effects.
And, as in all planet-search surveys, we must implement efficient methods 
for rejecting the many false alarms
that appear in the photometric light curves.
These false alarms result almost entirely from eclipsing systems involving
2 or more stars, including grazing eclipsing binaries, small stars
transiting large ones, and eclipsing binaries diluted by the light of
a third star.
For bright-star searches, these imposters can outnumber true planetary
transits by an order of magnitude \citep{bro03}.
Because of the diverse nature of the false alarm sources, several kinds
of follow-up observations are needed to reject them all \citep{alo04}.
We report here the first transiting extrasolar planet to be detected by
such a wide-field, bright-star survey.
We also describe the confirmation process in some detail, as an illustration
of the necessary steps in verifying that transits are caused by
an object of planetary and not stellar mass.

\section{Observations}

Our initial photometric observations leading to the detection of 
a planetary transit signature were conducted using the 3 telescopes
of the Trans-Atlantic Exoplanet Survey (TrES) network.
These telescopes (STARE, located on Tenerife in the Canary Islands,
PSST, located at Lowell Observatory, Arizona, and Sleuth, 
located at Mt. Palomar,
California)
\footnote{See also http://www.hao.ucar.edu/public/research/stare/stare.html and
http://www.astro.caltech.edu/$\sim$ftod/sleuth.html}
are being described individually elsewhere \citep{dun04, bro04}.
Briefly, all 3 are small-aperture (10 cm), wide-field (6$^{\circ}$),
CCD-based systems with spatial resolution of about 11$\arcsec$ per pixel.
They usually observe in red light (roughly Johnson $R$ for STARE and PSST,
Sloan $r$ for Sleuth), and they operate in coordination, observing the
same field in the sky continuously (or as nearly as possible) for
typically 2-month intervals.
The observing cadence at each site is roughly one image 
every 2 minutes, and the resulting time series are later binned
to 9-minute time resolution.
Recent adoption of an image-subtraction algorithm (based on
\citet{ala00}) yields photometric
precision of better than 2 mmag for the brightest non-saturating
stars ($R \simeq 8$), and better than 10 mmag
for $R \leq 12.5$.

We designate the planet described herein as TrES-1, the first confirmed
planet detected using the TrES network;
we refer to the parent star by the same name, since the distinction
between planet and star will be clear from context.
The star's coordinates, observed characteristics, and index numbers 
from various full-sky
catalogs are given in Table 1.
The $V$ and $B-V$ values come from differential photometry relative to 32 
stars with $B$ and $V$ data in SIMBAD; $R_{\rm J}$ was obtained 
from observations of Landolt's standards \citep{Land92}, and the $JHK$ values 
are from the 2MASS catalog.
The field containing this star was observed by 2 sites (STARE and PSST)
during
the summer of 2003,  
with STARE obtaining 49 good nights of observations and PSST 25.
The Sleuth telescope was still under development
at that time, and so did not observe this field.

The top panel of
Figure 1 shows the near-transit portion of the light curve of TrES-1,
folded with a period of 3.030065 days.
This curve is a superposition of 4 full transits and 2 partial ones, 
all observed with the
STARE telescope.
Even though the PSST telescope obtained 25 nights of good observations 
on the field, it observed no transits of TrES-1 in 2003.
This circumstance arose because the orbital period is very nearly an
integral number of days, so that for long intervals, transits can
be observed only from certain longitudes on the Earth.
Although data from PSST played no role in detecting the
transits, its data proved essential for a correct determination of
the orbital period:  we rejected several candidate periods because they
implied transit events that were not seen from the western US.
TrES-1 is thus a graphic demonstration of the utility of a
longitude-distributed network of transit-detection telescopes.

The $R$-band transit seen by TrES has a flat-bottomed shape, a depth
of 0.023 mag, and a total duration of about 3 hr.
These characteristics are consistent with expectations for a Jupiter-sized
planet crossing a cool dwarf star, but both experience 
\citep{lat03, char04} and
theory \citep{bro03} show that they are more likely to result from an
eclipsing stellar system.
Multiple star systems, in which the eclipsing binary component contributes 
only a small fraction of the total light, are particularly insidious.
Thus, TrES-1 was one of 16 stars that displayed transit-like
events among the 12000 stars we monitored in its surrounding field.
We therefore began an extensive program of observations with larger
telescopes, to determine whether the eclipses actually result from
a body of planetary mass.

From Table 1, the $J-K$ color of 0.48 suggests a star with spectral type
of late G or early K.
Digitized Sky Survey images show no bright neighbors 
within the 20$\arcsec$ radius
of a STARE stellar image, and adaptive-optics $H$- 
and $K$-band imaging with the 
William Herschel Telescope
showed no companion within 2 mag in brightness, farther than 0.3$\arcsec$
from the primary star. 
With its observed $V$ magnitude of 11.79, and ignoring interstellar
extinction, the implied distance to TrES-1 is about 150 pc.
Combining this distance with the USNO-B1.0 proper motion of 47 mas y$^{-1}$
\citep{mon03}
gives a transverse velocity of 26 km s$^{-1}$, which is fairly typical for
low-mass field stars in the solar neighborhood.
Thus, the photometric and astrometric evidence tends to confirm that
most of the detected light comes from a nearby dwarf star.

We observed the star using the CfA Digital Speedometers \citep{lat92}
at 7 different epochs, giving coverage of the full orbital phase.
These instruments record 4.5 nm of spectrum centered on the Mg b lines,
with spectral resolution of about 8.5~km~s$^{-1}$.
For the 7 exposures spanning 60 days we determined a
mean velocity of $-$20.52 km~s$^{-1}$.  
The average internal error estimate and
actual velocity rms achieved were both 0.39 km s$^{-1}$, suggesting that any
companion orbiting with a 3.03-day period must have a mass smaller than 5
$M_{\rm Jup}$.
This conclusion is not firm, however, if there
is blending light from a third component. 
From comparisons of our observed spectra with synthetic spectra
calculated by J. Morse using Kurucz models
(Morse \& Kurucz, private communication), 
we estimate that TrES-1
has $T_{\rm eff} = 5250 \ \pm 200$ K, $\log(g) = 4.5 \ \pm 0.5$, 
$v\sin i \leq 5$ km s$^{-1}$,
and metallicity similar to that of the Sun.
The slow rotation is particularly significant, for several reasons:
it indicates that the star has not been spun up by tidal interactions
with a massive secondary, it forecloses some blending scenarios, and
it means that more precise radial velocity measurements can be
obtained fairly readily.

We also obtained
a moderate resolution echelle spectrum covering the entire visible wavelength
range, using the Palomar 1.5m telescope.
Based on the comparison of this spectrum with the spectral standards
of \citet{mon99}, we classify the star as K0V; it shows no sign of
a composite spectrum nor of other peculiarities.

Many multiple-star configurations involve components with different
colors, which cause the blended eclipses to have color-dependent
depths.
Moreover, the detailed shape of eclipse light curves provides two
independent estimates of the secondary's size, relative
to that of the primary star.
One of these estimates comes from the eclipse depth, and the other
from the duration of the eclipse's ingress and egress portions
\citep{bro01, sea03}.
Consistency between these estimates is an indication that blending with
light from a third star is not important.
We therefore obtained multicolor photometric observations of several
transits, using larger telescopes and a variety of filters.
At the IAC 80-cm telescope, we observed a partial transit (missing
the egress) with Johnson $V$ and $I$ filters;
at the University of Colorado Sommers-Bausch Observatory 61-cm telescope,
we observed a full transit with Johnson $B$ and $R$ filters;
at the CfA's Fred L. Whipple Observatory 1.2-m telescope, we
observed one full and one partial transit with Sloan $g$, $r$, and $z$ filters.
The PSST telescope also observed 4 transits in $R$ during the 2004 season.
\hskip-4pt \footnote
{The photometric and radial velocity data described in the text are available
at http://www.hao.ucar.edu/public/research/stare/data/TrES1.asc.}
Figure 1 displays all of these observations, along with a fit to a
model, which we shall discuss below.
The light curves show no evidence for color dependence of the transit
depth (beyond that expected from color-dependent stellar limb darkening),
and both the transit depth and the short ingress/egress times 
are consistent with transits by an object whose radius is a small fraction
(less than about 0.15) of the primary star's radius.

Detailed modeling of the light curves following \citet{tor04b}
was carried out in an attempt to explain the observations as the
result of blending with an eclipsing binary. We found all plausible fits 
to be inconsistent with constraints from the CfA spectroscopy.
We conclude that TrES-1 is not significantly blended with the light of
another star.

On the strength of the foregoing analysis we obtained precise radial
velocity measurements using the I$_2$ absorption cell and HIRES
spectrograph on the Keck I telescope. Eight observations were
collected over a period of 18 days in July 2004, providing good
coverage of critical phases. The data reduction involved modeling of
the temporal and spatial variations of the instrumental profile of the
spectrograph \citep{val95}, and is conceptually similar to
that described by \citet{but96}. Internal errors were computed
from the scatter of the velocities from the echelle orders containing
I$_2$ lines, and are typically 10-15 m s$^{-1}$.  Figure 2 shows the radial
velocity measurements, along with a fit to a sinusoidal variation that
is constrained to have the period and phase determined from the
photometric data.  This constrained fit matches the data well, and
yields a velocity semi-amplitude of $K = 115.2 \ \pm 6.2$ m s$^{-1}$. 
The rms residual of
the fit is 14 m s$^{-1}$, in good agreement with the average of the internal
errors.  Examination of the spectral line profiles in our Keck spectra
by means of the bisector spans \citep{tor04a} indicated
no significant asymmetries and no correlation with orbital phase, once
again ruling out a blend.

\section{Discussion}

For purposes of an initial estimate of the planetary mass and radius,
we assumed
TrES-1 to have $T_{\rm eff} = 5250$K and solar metallicity.
By comparing with the accurately-known mass and radius of $\alpha$ Cen B,
which has a similar $T_{\rm eff}$ but probably higher metallicity
\citep{egg04}, and correcting for the assumed metallicity difference
of $\delta {\rm [Fe/H]}=-0.2$ using evolutionary models by \citet{gir00},
we estimate a stellar mass
$M_{\rm s} = 0.88 M_\sun$ and a radius
$R_{\rm s} = 0.85 R_\sun$.
We took limb darkening relations from \citet{cla00} and from Claret
(private communication), for models
with solar metallicity, $\log(g) = 4.5$, and $T_{\rm eff} = 5250$ K.
We assign (somewhat arbitrarily) an 
uncertainty of $\pm 0.07 M_{\sun}$ to $M_{\rm s}$.
We also take
$0.80 R_{\sun} \ \leq R_{\rm s} \leq \ 0.95 R_{\sun}$, 
since adequate fits to the photometry cannot
be obtained for stellar radii outside this range.

Using the approximate orbital period and the
constraints and assumptions just described, we estimated
the orbital semimajor axis $a$ and planetary mass $M_{\rm p}$ from 
the observed stellar reflex velocity and Kepler's
laws.
We then performed a minimum-$\chi^2$ fit to all of the photometry
(with errors estimated from the internal scatter of the input data,
taken when possible from the out-of-transit data only),
to obtain estimates for the planetary radius $R_{\rm p}$ and orbital
inclination $i$, and refined estimates for the epoch of transit center
$T_{\rm c}$ and for the orbital period $P$.
Our best estimates of the planet's orbital and physical parameters
are given in Table 2, and the solid curves in Figs 1 and 2 show the
fitted photometric and radial velocity variations overplotted on the
data.

The error estimates given in Table 2 include errors that follow
from our uncertainty in the radius and mass of the parent star
(which is assumed to be a main-sequence object),
as indicated in Table 1.
These uncertainties (especially in $R_{\rm s}$) dominate
errors in the photometry as regards estimates of $R_{\rm p}$ and $i$.
If the stellar radius and mass were known accurately, 
the uncertainties in $R_{\rm p}$ and
in $i$ would be smaller by about a factor of 10.
Contrariwise, if the star is actually a subgiant (photometric constraints
notwithstanding), $R_{\rm p}$ could exceed the upper limit in Table 2.
The error in $M_{\rm p}$ arises about equally from the radial velocity
measurement precision and from our
uncertainty in $M_{\rm s}$.

The mass, orbital radius, and radiative equilibrium temperature of
TrES-1 are quite similar to those of HD 209458b,
yet the former planet's radius is about 20\% smaller.
Indeed, as shown in Figure 3,
the radius of TrES-1 is more similar to those of the OGLE
planets, and it closely matches current models for irradiated planets
without internal energy sources \citep{cha04, bur04}.
This discrepancy between the radii of HD 209458b and TrES-1 reinforces
suspicion that HD 209458b has an anomalously large radius.

The confrontation between theory and observation for this object would
be facilitated if the stellar radius and (to a lesser degree)
mass could be better constrained.
We are undertaking a careful study of the Keck spectra of TrES-1, and
we will report improved estimates of the stellar parameters derived from
them in a later paper.
In the long run, however, a better approach is to obtain improved observations.
With spaceborne photometry, one can achieve low enough noise to
fit for both the planetary and the stellar radius \citep{bro01}.
Though one still  requires a guess for $M_{\rm s}$,  the derived planetary properties
are much less sensitive to this parameter than they are to $R_{\rm s}$.
Similarly, an accurate parallax measurement would imply a useful constraint
on $R_{\rm s}$.
Thus, TrES-1 may be an attractive target for either
ground- or space-based interferometric astrometry,
since it is relatively bright ($K = 9.8$), and it has several neighbors
of similar brightness within a radius of a few arcminutes.

\acknowledgments

We thank the students and staff of many observatories for their assistance
in obtaining the data needed for this study.
At FLWO, we thank Perry Berlind, Mike Calkins, and Gil Esquerdo.
At the Oak Ridge Observatory, Joe Caruso, Robert Davis,
and Joe Zajac.
At Iza\~na/Tenerife, Cristina Abajas, Luis Chinarro, 
Cristina D\'\i az, Sergio Fern\'andez, Santiago L\'opez, and Antonio 
Pimienta.
At SBO, Christine Predaina, Lesley Cook, Adam Ceranski, Keith Gleason,
and John Stocke.
We also thank A. Claret for computing limb-darkening coefficients for the
Sloan filter set, and J. Morse for helping with models for 
stellar classification.
Partial support for some of this work was provided through NASA's
Kepler Project, W. Borucki, PI.
The IAC-80 telescope is 
operated by the Instituto de Astrof\'\i sica de Canarias in its 
Observatorio del Teide.
This publication makes use of data products from the 
Two Micron All Sky Survey, which is a joint project 
of the University of Massachusetts and the Infrared
Processing and Analysis Center/California Institute of Technology, 
funded by the National Aeronautics and Space Administration 
and the National Science
Foundation.
Some of the data presented herein were obtained at the W. M. Keck Observatory, 
which is operated as a scientific partnership among the
California Institute of Technology, the University of California, 
and the National Aeronautics and Space Administration. 
The Observatory was made possible by the
generous financial support of the W. M. Keck Foundation.

\clearpage

\figcaption[fig3.eps]
{Time series photometry used in estimating the radius and inclination
of TrES-1, plotted against heliocentric time modulo the orbital
period from Table 2.
The telescope and filter bandpass used are indicated on each plot.
Each set of observations is overplotted with the predicted light curve
for that color, given the parameters in Table 2.
}

\figcaption[fig4.eps]
{Radial velocity observations of TrES-1, overplotted with
the best-fit orbit.}

\figcaption[fig5.eps]
{Radii of transiting extrasolar planets plotted against their masses.
Dashed curves are lines of constant density.
Data are from \citet{bro01} for HD 209458b, \citet{tor04a} for OGLE-TR-56,
\citet{bou04} and \citet{kon04} for OGLE-TR-113, \citet{mou04} for OGLE-TR-132, and the
present work for TrES-1.}
\clearpage

\begin{deluxetable}{lll}
\tabletypesize{\small}
\tablecaption{TrES-1 Parent Star}
\startdata
RA = 19:04:09.8 (J2000)& & Dec = +36:37:57  (J2000) \\
$R$ = 11.34 & $V$ = 11.79 & $B-V$ = 0.78\\
$J$ = 10.294 &$J-H$ = 0.407 &$H-K$ = 0.068 \\ 
Spectrum = K0V &$M_{\rm s} = 0.88 \  \pm 0.07 M_{\sun}$&$R_{\rm s} = 
0.85^{+0.10}_{-0.05} R_{\sun}$ \\ 
GSC & 02652-01324 & \\
2MASS & 19040985+3637574& \\
\enddata
\end{deluxetable}

\begin{deluxetable}{ll}
\tabletypesize{\small}
\tablecaption{TrES-1 Planet}
\startdata
Orbital Parameters & \\
$P \ = 3.030065\  \pm 8\times 10^{-6}$ d & $T_{\rm c} \ = \ 2453186.8060 \ \pm 0.0002$ (HJD)\\
$a \ = \ 0.0393 \ \pm 0.0011$ AU & $i \ = \ {88.5^{\circ}} ^{+1.5}_{-2.2}$ \\
$K \ = \ 115.2 \ \pm \ 6.2$ m s$^{-1}$ & \\
 & \\
Physical Parameters & \\
$M_{\rm p} \ = \ 0.75 \ \pm 0.07 \ M_{\rm Jup}$ & $R_{\rm p} \ = \ 1.08^{+0.18}_{-0.04}  \ R_{\rm Jup}$ \\
 & $R_{\rm p}/R_{\rm s} \ = \ 0.130^{+0.009}_{-0.003}$ \\
\enddata
\end{deluxetable}

\clearpage

\begin{figure}
\plotone{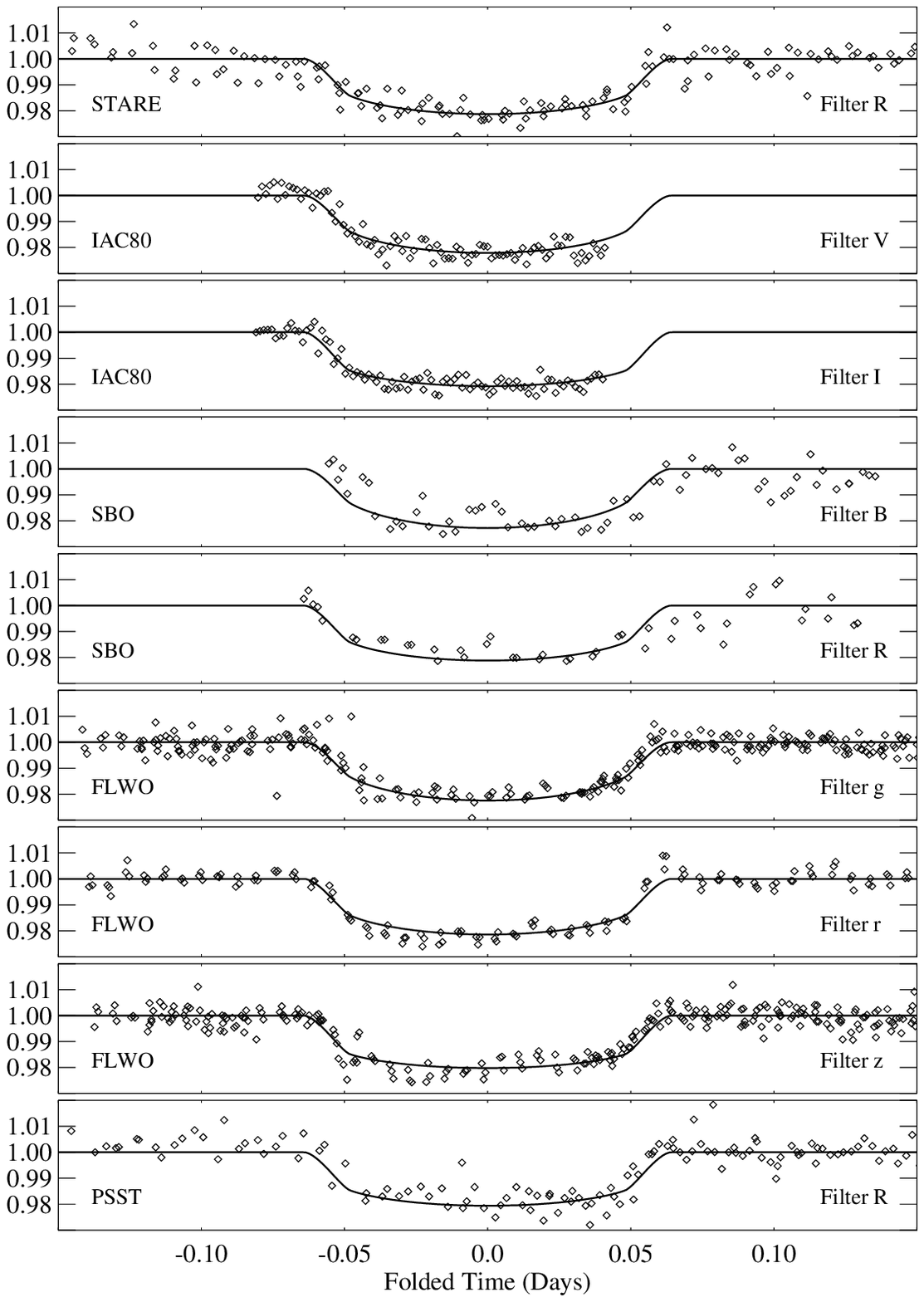}
\end{figure}

\begin{figure}
\plotone{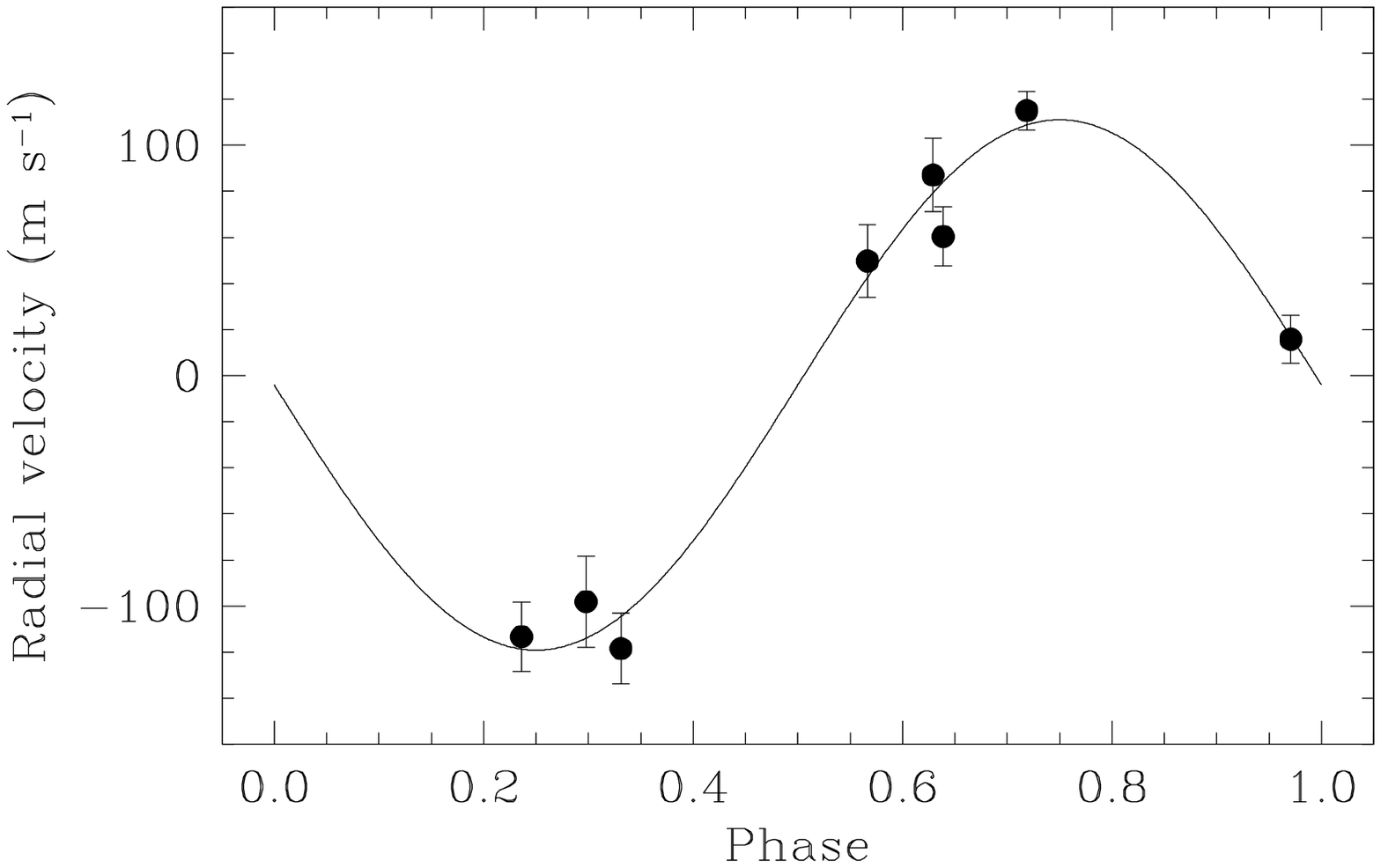}
\end{figure}

\begin{figure}
\plotone{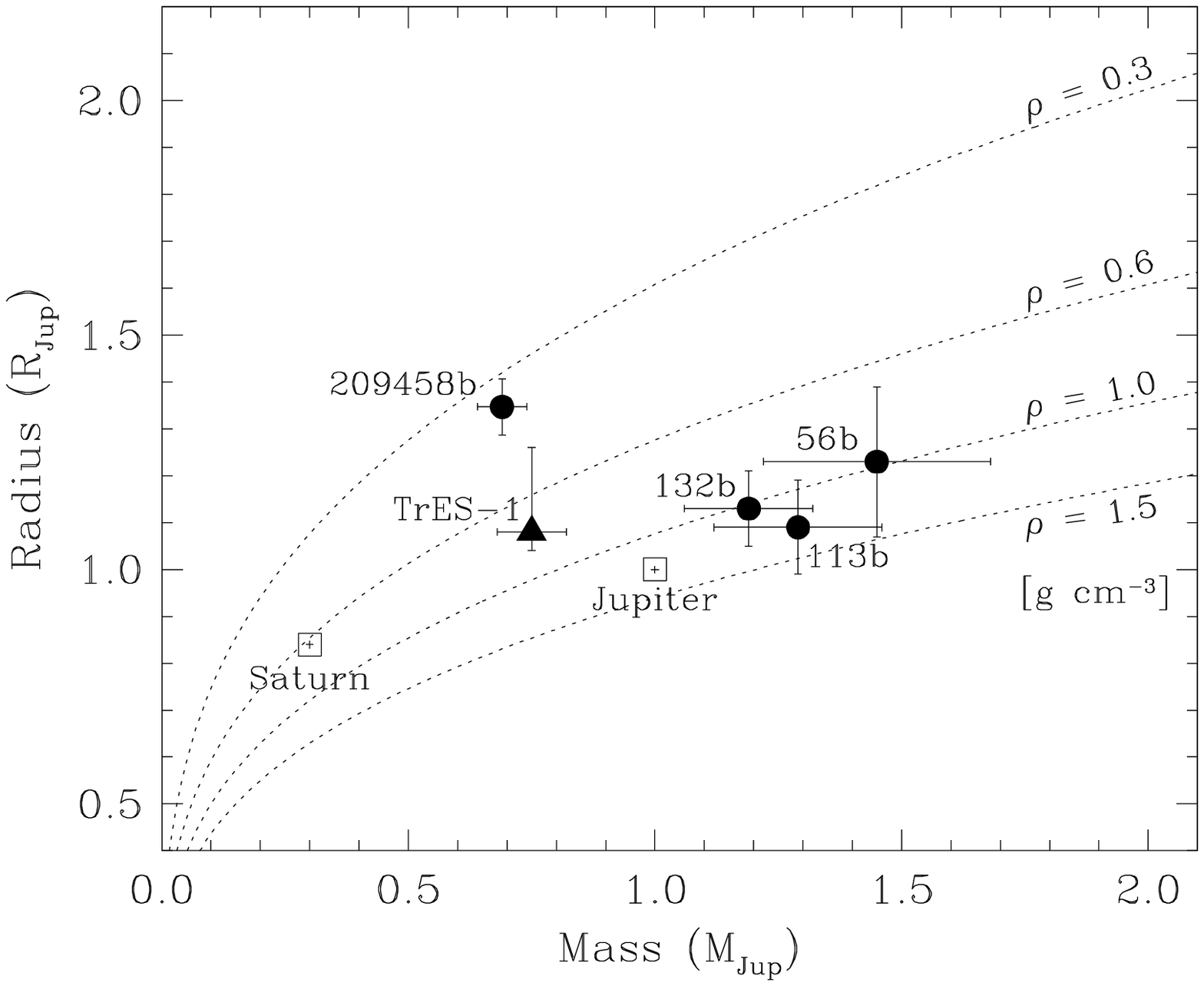}
\end{figure}

\end{document}